\documentclass[preprint,numbers]{sigplanconf_mod}

\usepackage{amsmath}
\usepackage{hyperref}

\begin{document}

\setlength{\pdfpageheight}{\paperheight}
\setlength{\pdfpagewidth}{\paperwidth}

\conferenceinfo{CONF 'yy}{Month d--d, 20yy, City, ST, Country}
\copyrightyear{20yy}
\copyrightdata{978-1-nnnn-nnnn-n/yy/mm}
\copyrightdoi{nnnnnnn.nnnnnnn}

% Uncomment the publication rights you want to use.
%\publicationrights{transferred}
%\publicationrights{licensed}     % this is the default
%\publicationrights{author-pays}

\titlebanner{}        % These are ignored unless
\preprintfooter{Pure Dataflow Matrix Machines - Version 3.0}   % 'preprint' option specified.

\title{Notes on Pure Dataflow Matrix Machines}
\subtitle{Programming with Self-referential Matrix Transformations}

\authorinfo{Michael Bukatin}
           { HERE North America LLC\\
              Burlington, Massachusetts, USA}
           {bukatin@cs.brandeis.edu}
\authorinfo{Steve Matthews}
           { Department of Computer Science\\
              University of Warwick\\
              Coventry, UK}
           {Steve.Matthews@warwick.ac.uk}
\authorinfo{Andrey Radul}
           { Project Fluid\\
             Cambridge, Massachusetts, USA}
           {aor021@gmail.com}

\maketitle

\begin{abstract}
Dataflow matrix machines are self-referential generalized recurrent neural nets.
The self-referential mechanism is provided via a stream of matrices defining
the connectivity and weights of the network in question. A natural question is:
what should play the role of untyped lambda-calculus for this programming architecture?
The proposed answer is a discipline of  programming with only one kind of streams, namely
the streams of appropriately shaped matrices. This yields {\em pure dataflow
matrix machines} which are networks of transformers of streams of matrices
capable of defining a pure dataflow matrix machine.
\end{abstract}

\category{D}{3}{2}[Data-flow languages]

% general terms are not compulsory anymore,
% you may leave them out
\terms
higher-order programming, dataflow

\keywords
continuous deformation of software, self-referential software

\section{Introduction}

The purpose of these notes is to contribute to the theoretical understanding of
dataflow matrix machines.

Dataflow matrix machines (DMMs) arise in the context of synchronous dataflow programming with
linear streams, i.e. streams equipped with an operation of taking a linear combinations of several streams~\cite{MBukatinMatthewsLinear}.

This is a new general-purpose programming architecture with interesting properties. One of these properties
is that large classes of programs are parametrized by matrices of numbers. In this aspect DMMs
are similar to recurrent neural nets and, in fact, they can be considered to be a very powerful
generalization of recurrent neural nets~\cite{MBukatinMatthewsRadulDMM,MBukatinMatthewsRadulPatterns}. 

Just like recurrent neural nets, DMMs are essentially ``two-stroke engines". On the ``up movement"
the built-in neuron transformations compute the next elements of the streams associated with the neuron outputs
from the streams associated with neuron inputs. This computation is local to the neuron in question
and is generally nonlinear. On the ``down movement", the next elements of the streams associated with 
all neuron inputs are computed from the streams associated with all neuron outputs
using the matrix controlling the DMM. This computation is linear and is potentially quite global, as any neuron output in the net
can contribute to any neuron input in the net.

DMMs described in the literature are heavily typed. One normally defines a finite collection of
allowed kinds of linear streams, and a finite collection of allowed types of neurons. These two collections are called the
{\em DMM signature}. One considers a particular fixed signature.
Then one assumes the address space accommodating a countable number
of neurons of each type, and then a DMM is determined by a countable-sized matrix of
connectivity weights (one normally assumes that only a finite number of those weights
are non-zero at any given moment of time).

In particular, DMMs can be equipped with powerful reflection facilities. Include
in the signature
the kind of streams of matrices shaped in such a fashion as to be capable of describing
a DMM over this signature. Then designate a particular neuron, \texttt{Self}, working as
an accumulator of matrices of this shape, and agree that the most recent output of this neuron will be
used at the ``down movement" of each step as the matrix controlling the calculations of all neuron inputs from
all neuron outputs.

\subsection{One Kind of Streams}\label{one_stream}

DMMs seem to be a powerful programming platform. In particular, it is convenient
to manually write general-purpose software as DMMs. At the same time the options
to automatically synthesize DMMs by synthesizing the matrices in question are available.

However, DMMs are a bit too unwieldy for a theoretical investigation.

From the theoretical viewpoint, it is inconvenient that there are many kinds of streams.
It is also inconvenient that one needs to fix a signature, and that the parametrization by matrices
is valid only for this fixed signature.

So a question naturally arises: {\em What would be the equivalent of untyped lambda-calculus for dataflow matrix machines?}

One of the principles of untyped lambda-calculus: one data type is enough, namely the type
of programs. All data can be expressed as programs.

The equivalent of this principle for DMMs would be to have only one kind of streams: streams of matrices,
where a matrix is so shaped as to be able to define a DMM which would be a network of transformers of streams of matrices
(see Section~\ref{address_space} for details).

Instead of string rewriting, a number of streams of matrices are unfolding in time in this approach.

So all data are to be expressed as countably-infinite matrices of numbers under this approach (see Section~\ref{as_matrices}), just like
all data must be expressed as lambda-terms in the untyped lambda-calculus.

\subsection{One Signature}\label{one_signature}

Choosing a fixed selection of types of neurons seems too difficult at the moment. For the time being
we would like to retain the ability to add arbitrary types of neurons to our DMMs. 

So instead of selecting a fixed canonical signature, we assume that there is an underlying language allowing to describe countable collection
of neuron types in such a fashion that all neuron types of interest can be expressed in that language.

Then assume that all neuron types described by all neuron type expressions in the underlying
language are in the signature. Assume that our address space is structured in such a way
as to accommodate countable number of neurons for each type of neurons (see Section~\ref{address_space}). Since we have
a countable collection of expressions describing neuron types, our overall collection of neurons is
still countable, and the matrix describing the rules to recompute neuron inputs from the neuron outputs
is still countable.

So, now we have a parametrization by countable matrices of numbers across all DMMs, and not just across DMMs with a particular
fixed signature.

\subsection{Accumulators Revised}\label{accum_revised_small}

The notion of accumulator plays a key role in a number of DMM constructions including the reflection facility \texttt{Self}.

The most standard version is a neuron performing an identity transform of its vector input, $x$, to
its vector output, $y$, of the same kind. One sets the weight of the recurrent connection from $y$ to $x$ to 1,
and then the neuron accumulates contributions of other neurons connected to $x$ with nonzero weights.

So, at each step the accumulator neuron in effect performs $v := v + \Delta v$ operation. However, it is somewhat
of abuse of the system of kinds of streams to consider $v$ and $\Delta v$ as belonging to the same space, and we'll see evidence that
to do so is a suboptimal convention later in the paper.

So, what we do, first of all, is that we equip the accumulator neuron with another input, where $\Delta v$ is
collected. Then the body of the neuron computes the sum of $v$ and $\Delta v$, instead of just
performing the identity transform (see Section~\ref{accum_revised} for more details).

In the situations, where one has multiple kinds of linear streams, one would often want to assign different
kinds to $v$ and to $\Delta v$ (although in other situations one would still use the same kind for the both of them, effectively considering
$\Delta v$ to be $0+\Delta v$).

\subsection{Structure of the Paper}

In Section~\ref{continuous_models} we discuss continuous models of computation and their higher-order aspects.
In Section~\ref{higher_order} we  juxtapose string rewriting with stream-based approaches to higher-order programming.
In Section~\ref{address_space} we discuss the language of indexes of the network matrix and how to accommodate countable
number of neuron types within one signature. In Section~\ref{as_matrices} we discuss representation of constants
and vectors as matrices.

Section~\ref{accum_revised} provides two examples where it is natural to split the accumulator input into
$v$ and $\Delta v$. One such example comes from the neuron \texttt{Self} controlling the network matrix.
Another example (Section~\ref{warmus_numbers}) is more involved and requires us to revisit domain theory in the context of
linear models of computation. This is a bitopological setting, more specifically, bi-continuous domains, allowing for both monotonic and
anti-monotonic inference, and this is the setting where approximations spaces tend to become embedded
into vector spaces, which is where the connection with linear models of computation comes into play.

\section{Continuous Models of Computation}\label{continuous_models}

The history of continuous models of computation is actually quite long. Where the progress was more
limited was in making higher-order constructions continuous, in particular, in making spaces of programs
continuous. Denotationally, the continuous domains representing the meaning of programs are common.
But operationally, we tend to fall back onto discrete schemas. 

Dataflow matrix machines are seeking to
change that and to provide programming facilities using continuous programs and continuous deformations
of programs on the level of operational semantics and of implementation. This can be done both for
discrete time and discrete index spaces (countably-sized matrices of computational elements), and, potentially, for continuous time and continuous
index spaces for computational elements.

The oldest electronic continuous platform is electronic analog computers. The analog program itself, however, is very discrete, 
because this kind of machine has a number of single-contact sockets and for every pair of such sockets there is
an option to connect them via a patch cord, or not to connect them.

Among dataflow architectures oriented towards handling the streams of continuous data one might mention
LabVIEW~\cite{WJohnstonHannaMillar} and Pure Data (e.g.~\cite{AFarnell}). In both cases, the programs themselves
are quite discrete.

The computational platform which should be discussed in more details in this context is recurrent neural networks.
Turing universality of recurrent neural networks is known for at least 30 years~\cite{JPollack,HSiegelmannSontag}.

However, together with many other useful and elegant Turing-universal computational systems, recurrent neural networks
do not constitute a convenient general-purpose programming platform, but belong to the class of {\em esoteric programming
languages} (see~\cite{MBukatinMatthewsRadulDMM,MBukatinMatthewsRadulPatterns} for detailed discussion of that).

Interestingly enough, whether recurrent neural networks understood as programs are discrete or continuous depends
on how one approaches the representation of network topology. If one treats the network connectivity as a graph, and thinks
about this graph as a discrete data structure, then recurrent neural networks themselves are discrete.

If one states instead that the network connectivity is always the complete graph, and that the topology is defined by
some of the weights being zeros, then recurrent neural networks themselves are continuous.

The most frequent case is borderline. One considers a recurrent neural net to be defined by the matrix of weights,
and therefore to be continuous, however there are auxiliary discrete structures, e.g. the matrix of weights is
often a sparse matrix, and so a dictionary of nonzero weights comes into play. Also a language used in describing
the network or its implementation comes into play as an auxiliary discrete structure.

Dataflow matrix machines belong to this borderline case. In particular, the use of sparse matrices is inevitable,
because the matrices in question are countable-sized matrices with finite number of nonzero elements.

\section{Higher-order Programming: String Rewriting vs. Stream-based Approach}\label{higher_order}

There are several approaches to higher-order stream-based programming. The most popular
approach starts with standard higher-order functional programming and focuses on integrating
stream-based programming into that standard paradigm. The theoretical underpinning of this
approach is lambda-calculus and string rewriting (e.g.~\cite{NGoodmanMansinghkaRoyBonawitzTenenbaum}).

The dataflow community produced purely stream-based approaches to higher-order programming.
One of those approaches which should be mentioned is an approach based on
multidimensional streams~\cite{WWadgeHigherOrderLucid}.

The approach which we adopt in this paper is based on the notion of streams of programs.
An early work which should be mentioned in connection with this approach is~\cite{SMatthews}.
An argument in favor of this approach for programming with linear streams was presented in Section 3 of~\cite{MBukatinMatthewsLinear}.

Among recent papers exploring various aspects of the approach based on the notion of streams of programs are~\cite{NKrishnaswami,MBukatinMatthewsRadulDMM,MBukatinMatthewsRadulPatterns}.
One of the goals of the present paper is to show that this approach can play the role in
synchronous dataflow programming with linear streams comparable to the role played by untyped lambda-calculus in functional programming.

\section{DMM Address Space: Language of Indices}\label{address_space}

When one has a countable-sized matrix, it is often more convenient to index its rows and columns by finite strings
over a fixed finite alphabet than by numbers. There is no principal difference, but this choice discourages focusing on an arbitrary chosen order,
and encourages semantically meaningful names for the indices.

Here we explain how the construction promised in Section~\ref{one_signature} works.

\subsection{Neuron Types}\label{neuron_types}

Define the notion of a type of neurons following the outline presented in Section 3.1 of~\cite{MBukatinMatthewsRadulDMM}
for multiple kinds of linear streams.
We only have one kind of linear streams in the present paper, so the definition is simplified. 
A neuron type consists of a non-negative integer input arity $M$, a positive integer output arity $N$,
and a transformation describing how to map $M$ input streams of matrices into $N$
output streams of matrices.

Namely associate with the neuron type in question a transformation $F$ taking as inputs $M$ streams of length $t-1$ and producing as outputs $N$ streams
of length $t$ for integer time $t>0$. Require
the obvious prefix condition that when $F$ is applied to streams of length $t>0$, the first $t$ elements of the output streams of length $t+1$ are the elements
which $F$ produces when applied to the prefixes of length $t-1$ of the input streams.
The most typical situation is when for $t>1$ the $t$'s elements of the output streams are produced solely
on the basis of elements number $t-1$ of the input streams, but our definition also allows neurons to
accumulate unlimited history, if necessary.

\subsection{Language $L_T$}

In this section we are going to use several alphabets. Assume that the following special symbols don't belong to any of the
other alphabets: $\Sigma_* = \{$\texttt{\textbackslash\%@}$\}$.

Assume that there is a language $L_T$ over alphabet $\Sigma_T$, such that finite strings from $L_T^T \subset L_T$ describe
all neuron types of interest. Call a string $t$ the name of the neuron type it defines (we are not worried about uniqueness of names for a type here).
Assume that the input arity of the type in question is $M_t$ and the output arity of the type in question is $N_t$.
That for every integer $i$ such that $0< i \leq M_t$ associate field name $I_{ti}$ from $L_T$ and for every integer
$j$ such that $0 < j \leq N_t$ associate field name $O_{tj}$ from $L_T$, so that $i_1 \neq i_2$ implies $I_{ti_1} \neq I_{ti_2}$ and
$j_1 \neq j_2$ implies $O_{tj_1} \neq O_{tj_2}$.

Also assume that there is an alphabet $\Sigma_0$ with more than one letter in it and any finite string $s$ over $\Sigma_0$ is a valid {\em simple name}.

\subsection{Language of Indices}

The following convention describes the address space for a countable number of neurons for each
of the countable number of neuron types of interest. The indexes are expressed by strings over the alphabet $\Sigma_* \cup \Sigma_T \cup \Sigma_0$.
 For any name of neuron type $t \in L_T^T$ and for any simple name $s$,
the concatenation $t +$ \texttt{@} $+ s$ is a name of a neuron.

For any field name $I_{ti}$, 
the concatenation $t +$ \texttt{@} $+I_{ti}+$ \texttt{\textbackslash} $+ s$ is the name of the corresponding neuron input.
For any field name $O_{tj}$,
the concatenation $t +$ \texttt{@} $+O_{tj}+$ \texttt{\%} $+ s$ is the name of the corresponding neuron output. 
For every such
pair of indices,\linebreak  $(t_1 +$ \texttt{@} $+I_{t_1i}+$ \texttt{\textbackslash} $+ s_1, t_2 +$ \texttt{@} $+O_{t_2j}+$ \texttt{\%} $+ s_2)$, there is
a matrix element in our matrices under consideration.

To summarize: in this approach the class of pure dataflow matrix machines is implicitly parametrized
by a sufficiently universal language $L_T$ describing all types of neurons  taken to be of potential interest
together with their associated built-in
stream transformations.

For details of DMM functioning see Sections 3.3 and 3.4 of~\cite{MBukatinMatthewsRadulDMM}.

\section{Constants and Vectors as Matrices}\label{as_matrices}

To implement the program outlined in Section~\ref{one_stream} one needs to express
the most important linear streams, such as streams of numbers (scalars), streams of matrix rows and
streams of matrix columns, and other frequently used streams of vectors as
streams of matrices. As indicated in~\cite{MBukatinMatthewsRadulDMM,MBukatinMatthewsRadulPatterns},
one of the key uses of scalars and also of matrix rows and columns is their use
as multiplicative masks. 

The ability to use scalars as multiplicative masks needs to be preserved when those
scalars are represented by matrices. For example, if we have a neuron which takes an input stream of scalars $a$
and an input stream of matrices $M$, and produces an output stream of matrices $a * M$, then we still
need to be able to reproduce this functionality when scalars $a$ are represented by matrices of the same
shape as matrix $M$.

The most straightforward way to do this is to have a neuron which takes two input streams
of matrices and performs their element-wise multiplication (Hadamard product, sometimes also called the Schur product).
If we chose the Hadamard product as our main bilinear operation on matrices, then the
scalar $x$ must be represented by the matrix all elements of which are equal to $x$.

\subsection{Matrices Admitting Finite Descriptions}

One particular feature of this approach is that we can no longer limit ourselves by matrices containing
finite number of non-zero elements, but we also need at least some infinite matrices admitting
finite descriptions. 

This means that one needs a convention of what should be done in case of incorrect operations, such as
taking a scalar product of two infinite vectors of all ones (or adding a matrix consisting of all ones to \texttt{Self}).
It seems likely that the technically easiest convention in such cases would be to output zeros (or to reset the network matrix to all zeros).

On the other hand, it is of interest to consider and study the limits of sequences of finitely describable matrices,
and a network might be computing such a limit when $t \rightarrow \infty$.

\subsection{Representing Matrix Rows and Columns as Matrices}

Streams of matrix rows and streams of matrix columns also play important roles in~\cite{MBukatinMatthewsRadulDMM,MBukatinMatthewsRadulPatterns}.
Represent element $y$ of a row by the corresponding matrix column all elements of which equal $y$.
Represent element $z$ of a column by the corresponding matrix row all elements of which equal $z$.

Hence, rows are represented by matrices with equal values along each column, and columns are
represented by matrices with equal values along each row.

Given matrix row $\alpha$, denote by $(^{\uparrow} \alpha)$ its representation as a matrix. Given matrix column $\beta$,
denote by $(\beta^{\rightarrow})$ its representation as a matrix. Given scalar $x$, denote by $(^{\uparrow} x ^{\rightarrow})$
its representation as a matrix.

Respecting the MATLAB convention to denote the Hadamard product by \texttt{.*}, we denote the Hadamard product of two
matrices by $A \dot{*} B$, while omitting the infix for matrix multiplication, $A^T B$ or $A  B^T$. 

Note that because
matrix rows correspond to neuron inputs and matrix columns correspond to neuron outputs, one should always think
about these matrices as rectangular, and not as square matrices, so the transposition is always needed when performing the
standard matrix multiplication on these matrices. 

In~\cite{MBukatinMatthewsRadulDMM} a standard matrix update operation generalized from several natural examples
is proposed. Given a row $\alpha$, two columns $\beta$ and $\gamma$ (with the constraint that both $\beta$ and $\gamma$ have finite
number of nonzero elements), the matrix is
updated by the formula $a_{ij} := a_{ij} + \gamma_i * \alpha_j * \sum_k \beta_k a_{kj}$.

In terms of matrix representations what gets added to the network matrix $A$ is   $(\gamma^{\rightarrow}) \dot{*} (^{\uparrow} \alpha) \dot{*} ( ^{\uparrow} (\beta^T A))$.

In Section 4 of~\cite{MBukatinMatthewsRadulPatterns} matrix rows and columns are used for subgraph selection.
Consider a subset of neurons, and take $\alpha$ to be a row with values 1 at the positions corresponding to the neuron outputs of
the subset in question and zeros elsewhere, and take $\beta$ to be a column with values 1 at the positions corresponding to the neuron inputs of
the subset in question and zeros elsewhere. Denote the element-wise matrix maximum as $A \dot{\sqcup} B$.

The overall connectivity of the subgraph in question is expressed by the matrix
$((^{\uparrow} \alpha) \dot{\sqcup} (\beta^{\rightarrow})) \dot{*} A$, while the internal connectivity of this subgraph is
$(^{\uparrow} \alpha) \dot{*} (\beta^{\rightarrow}) \dot{*} A$.

\subsection{Other Vectors as Matrices.}

The most straightforward way to represent other finite-dimensional vectors or
countable-dimensional vectors with finite number of nonzero elements in this setup
is to represent them as matrix rows as well. This means reserving a finite or
countable number of appropriately typed neurons to represent
coordinates.

For example, to describe vectors representing characters in the ``1-of-$N$" encoding
which is standard in neural nets~\cite{AKarpathy} one would need to reserve neurons to represent the letters
of the alphabet in question.

\section{Accumulators Revised}\label{accum_revised}

Here we continue the line of thought started in Section~\ref{accum_revised_small}.

We give a couple of examples illustrating why it is natural to have
separate inputs for $v$ and $\Delta v$ in an accumulator.

The main example is the neuron \texttt{Self} itself, producing the matrix controlling
the network on the output, and taking additive updates to that matrix on the input.
This is a countable-sized matrix with finite number of nonzero elements, so it has to be
represented as a sparse matrix, e.g. via a dictionary of nonzero elements. A typical situation is
that the additive update on each time step is small compared to the matrix itself
(more specifically, the update is typically small in the sense that the number of affected matrix elements
is small compared to the overall number of nonzero matrix elements).

So it does not make much sense to actually copy the output of \texttt{Self} to
its input of \texttt{Self} and perform the additive update there, which is what
should be done if the non-optimized definition of an accumulator with one input
is to be taken literally.

What should be done instead is that additive updates should be added together
at an input of \texttt{Self}, and then on the ``up movement" the \texttt{Self}
should add the sum of those updates to the matrix it accumulates.

So instead of hiding this logic as ``implementation details", it makes sense to split
the inputs of \texttt{Self} into $x$ (with the output of \texttt{Self} connected to
$x$ with weight 1, nothing else connected to $x$ with non-zero weight, and the copying
of the output of \texttt{Self} to $x$ being a no-op) and $\Delta x$ accumulating
the additive updates to \texttt{Self}.

\subsection{Partial Inconsistency Landscape and Warmus Numbers}\label{warmus_numbers}

Another example of why it is natural to have
separate inputs for $v$ and $\Delta v$ in an accumulator comes from
considering a scheme of computation with Warmus numbers.

We have to explain first what are Warmus numbers and why considering
them and a particular scheme of computation in question is natural in this context.

\subsubsection{Partial Inconsistency and Vector Semantics}

{\em In the presence of partial inconsistency approximation spaces tends to become embedded
into vector spaces.} One well-known example of this phenomenon is that if one allows negative values for
probabilities, then probabilistic powerdomain is embedded into the space of signed measures which is a
natural setting for denotational semantics of probabilistic programs~\cite{DKozen}.

\subsubsection{Warmus Numbers}

Another example involves algebraic extension of interval numbers with respect to addition.
Interval numbers don't form a group with respect to addition. However one can extend them with {\em pseudosegments}
$[b,a]$ with the contradictory property $a < b$. For example, [3,2] is a pseudosegment expressing an interval
number with the contradictory constraint that $x \geq 3$ and at the same time $x \leq 2$.
The so extended space of interval numbers is a group and a 2D vector space over reals.

The first discovery of this construction known to us was made by Mieczys\l{}aw Warmus~\cite{MWarmus}.
Since then it was rediscovered many times. For a rather extensive bibliography related to those rediscoveries
see~\cite{EPopova}.

\subsubsection{Partial Inconsistency Landscape}

There are a number of common motives which appear multiple times in various studies of partial inconsistency,
in particular, bilattices, bitopology, bicontinuous domains, facilities for non-monotonic and anti-monotonic inference,
order-reversing involutions, etc. Together, these motives serve as focal elements of the field of study which has been
named the {\em partial inconsistency landscape} in~\cite{MBukatinMatthewsLinear}.

In particular, the following situation is typical in the context of bitopological groups. The two topologies, $T$ and $T^{-1}$,
are group dual of each other  (that is, the group inverse induces a bijection between the respective systems of open sets), and the anti-monotonic group inverse is an order-reversing involution, which is a bicontinuous
map from $(X, T, T^{-1})$ to its bitopological dual, $(X, T^{-1}, T)$~\cite{SAndimaKoppermanNickolas}.

Because approximation domains tend to become embedded into vector spaces in this context, the setting of
bicontinuous domains~\cite{KKeimel} equipped with two Scott topologies which tend to be group dual of each
other seems to be natural for semantic studies of computations with linear streams.

\subsubsection{Computing with Warmus Numbers}

Section 4 of ~\cite{MBukatinMatthewsLinear} provides a detailed
overview of the partial inconsistency landscape, including the bitopological and
bilattice properties of Warmus numbers. It turns out that Warmus numbers play
a fundamental role in mathematics of partial inconsistency. In particular, Section 4.14 of that
paper proposes a schema of {\em computation via monotonic evolution punctuated by
order-reversing involutive steps}.

Computations with Warmus extension of interval numbers via monotonic
evolution punctuated by involutive steps  are a good example
of why the accumulators should have the asymmetry between $v$ and $\Delta v$.

If an accumulator neuron is to accumulate a monotonically evolving Warmus number by accepting additive updates
to that number, then the $\Delta x$ cannot be an arbitrary Warmus number, but it must be a
pseudosegment $[b,a]$, such that $a \leq 0\leq b$ (the case of $a=b=0$ is allowed). Given that there is
a constraint of this kind, it is natural to want to accumulate $\Delta x$ contributions at a separate input
on the ``down movement", and to let the accumulator enforce the constraint on the ``up movement"
(e.g. by ignoring requests for non-monotonic updates). Yet another input might be added to trigger involutive steps (an involutive
step in this context transforms $[d,c]$ into $[c,d]$).
Alternatively, requests for non-monotonic updates might trigger the involutions. Normally, the involution
would be triggered only if the accumulated number is already a pseudosegment, in which case the involution is
an anti-monotonic step.

\subsubsection{Open Problem: Bicontinuous Reflexive Domains}

Despite impressive progress in studies of bicontinuity and bitopology in the context of
partial inconsistency landscape~\cite{DKozen,AJungMoshier,KKeimel,JLawson}, the issues related
to reflexive domains and solutions of recursive domain equation in the context of bicontinuous domains
and vector semantics don't seem to be well understood.

Given that dataflow matrix machines equipped with 
self-referen-tial
facilities work directly on the level
of vector spaces, one would hope that the gap between operational and denotational descriptions
would be more narrow in this case than for more traditional situations such as untyped lambda-calculus.

\section{Conclusion}

Dataflow matrix machines work with arbitrary linear streams. In this paper, we focus on the case
of {\em pure dataflow matrix machines}, which work with the single kind of linear streams, namely 
the streams of matrices defining the connectivity patterns and weights in pure DMMs themselves.

This allows us to pinpoint the key difference between pure DMMs and
recurrent neural networks: instead of working with streams of numbers, pure dataflow
matrix machines work with streams of programs, with programs being represented as network
connectivity matrices.

%\appendix
%\section{Appendix Title}

%This is the text of the appendix, if you need one.

%\acks

%Acknowledgments, if needed.

% We recommend abbrvnat bibliography style.

\bibliographystyle{abbrvnat}

% The bibliography should be embedded for final submission.

%\bibliography{LinearModels3}

%\begin{thebibliography}{}
%\softraggedright

%\bibitem[Smith et~al.(2009)Smith, Jones]{smith02}
%P. Q. Smith, and X. Y. Jones. ...reference text...

%\end{thebibliography}

\appendix
\section{Monotonic Evolution by Additions: Warmus Numbers vs. Conventional Interval Numbers }\label{sec:interval_monotonic}

Consider a sequence $x \sqsubseteq  (x+x_1) \sqsubseteq  (x+x_1+x_2) \sqsubseteq \dots$ of elements,
which are monotonically increasing and are obtained by additive corrections from previous elements of
the sequence. 

If these are
conventional interval numbers, this situation is only possible for the trivial case of $0 = x_1 = x_2 = \dots$, as
addition cannot reduce the degree of imprecision (self-distance) for conventional interval numbers.
It is not possible to perform nontrivial monotonic evolution of conventional interval numbers by adding other
interval numbers to previous elements of the sequence in question.

For Warmus numbers, monotonic evolution by additive corrections is possible, provided that
every additive correction summand $x_i = [a_i, b_i]$ is a pseudo-segment anti-approximating zero:

$[0,0] \sqsubseteq [a_i,b_i]$, that is $b_i \leq 0 \leq a_i$.

\section{Rectifiers and Quasi-metrics}\label{sec:rectifiers_and_quasi_metrics}

Rectified linear unit (ReLU) is a neuron with the activation function $f(x) = max(0,x)$.

In the recent years, ReLU became the most popular neuron in the context of non-recurrent
deep networks. Whether it is equally good for recurrent networks remains to be seen.

The activation function $max(0,x)$ is an integral of the Heaviside step function. Lack of smoothness at 0 does not seem to
interfere with gradient methods used during neural net training.

Interestingly enough,  the standard quasi-metrics on reals associated with upper and lower
topologies on reals are closely related to ReLU:
$q_1(x,y) = f(x-y) = q_2(y,x)$.

\section{Linear and Bilinear Neurons in LSTM and Gated Recurrent Unit Networks}

Various schemas of recurrent networks with gates and memory were found to be useful 
in overcoming the problem of vanishing gradients in the training of recurrent neural networks, 
starting with LSTM in 1997 and now including a variety of other schemas.

For a convenient compact overview of LSTM, gated recurrent units networks, and related schemas see
Section 2 of~\cite{ZhouWuZhangZhou}.

The standard way to describe LSTM and gated recurrent unit networks is to think
about them as networks of sigmoid neurons augmented with external memory and gating mechanisms.

However, it is long understood (and is used in the present paper) that neurons with linear
activation functions can be used as accumulators to implement memory.

It is also known for at least 30 years that bilinear neurons (such as neurons multiplying
two inputs, each of those inputs accumulating linear combinations of output signals
of other neurons) can be used
to modulate signals via multiplicative masks (gates) and to implement
conditional constructions in this fashion~\cite{JPollack} (see also Section 1.3.2 of~\cite{MBukatinMatthewsRadulPatterns}).

Looking at the formulas for LTSM and gated recurrent unit networks in Table 1 of~\cite{ZhouWuZhangZhou}
one can observe that instead of thinking about these networks as networks of sigmoid neurons augmented with external memory and gating mechanisms,
one can describe them simply as recurrent neural networks built from sigmoid neurons, linear neurons, and
bilinear neurons, without any external mechanisms.

When LTSM and gated recurrent unit networks are built as recurrent neural networks from sigmoid neurons, 
linear neurons, and bilinear neurons, some weights are variable and subject to training, and some weights
are fixed as zeros or ones to establish a particular network topology.

\section{Lightweight Pure Dataflow Matrix Machines}

Pure dataflow matrix machines are countable-sized networks with a finite part of the network
being active at any given moment of time. They process streams of countable-sized matrices
with finite number of non-zero elements,

Sometimes it is convenient to consider the case of networks of finite size, with fixed number
of inputs, $M$, and fixed number of outputs, $N$. If we still would like those networks to
process streams of matrices describing network weights and topology, those matrices would
be finite rectangular matrices $M \times N$.

We call the resulting class of networks {\bf Lightweight Pure DMMs}.
If we work with reals of limited precision and consider fixed values of $M$ and $N$,
the resulting class is not Turing-universal, as its memory space is finite. However, it is often useful to consider
this class for didactic purposes, as both theoretical constructions and software
prototypes tend to be simpler in this case, while many computational effects
can already be illustrated in this generality.

\subsection{Dimension of the Network Operators}

The network has $N$ outputs, each of which is a matrix $M \times N$, hence the
overall dimension of the output space is $M \times N^{2}$.

The network has $M$ inputs, each of which is a matrix $M \times N$, hence the
overall dimension of the input space is $M^{2} \times N$.

So, overall the dimension of space of all possible linear operators from outputs
to inputs (which could potentially be used during the ``down movement") is
$M^{3} \times N^{3}$. However, our model actually uses matrices of the
dimension $M \times N$ during the ``down movement", so only a subspace of dimension $M \times N$ of
the overall space of all possible linear operators of the dimension $M^{3} \times N^{3}$
is allowed. The matrix is applied not to a vector of numbers, but to a
vector of $N$ matrices $M \times N$, and yields not a vector of numbers,
but a vector of $M$  matrices $M \times N$. This is what accounts for
factoring $M^{2} \times N^{2}$ dimension out.

\subsection{Software Prototypes}

We prototyped lightweight pure DMMs in Processing 2.2.1 in the {\tt Lightweight\_Pure\_DMMs} directory of
Project Fluid, which is our open source project dedicated to experiments with the computational architectures
based on linear streams~\cite{Fluid}.

For simplicity we used numbers to index rows and columns of the matrices, instead of using semantically
meaningful strings we recommend to use as indices for non-lightweight work.

In particular, we demonstrated during those experiments that it is enough to consider a set of several
constant update matrices together with our self-referential network update mechanism described in
the present paper to create oscillations of network weights and waves of network connectivity patterns.

\subsubsection{The {\tt aug\_26\_16\_experiment} directory}

Assume that the neuron {\tt Self} adds matrices $X^0$ and $X^1$ on the ``up movement" to
obtain matrix $Y^0$. Assume that at the starting moment $t=0$, $Y^0_{0,0} = 1$, $Y^0_{0,j} = 0$ for all $j \neq 0$,
$Y^0_{1,1} = 1$, $Y^0_{1,j} = 0$ for all $j \neq 1$.

Assume that $Y^1$ is a constant matrix, such that $Y^1_{0,j} = 0$ for all $j$,
$Y^1_{1.1} = -2$,  $Y^1_{1,j} = 0$ for all $j \neq 1$.

The network starts with a ``down movement".
After the first ``down movement", $X^0$ becomes a copy of $Y^0$, $X^1$ becomes a copy of $Y^1$,
and after the first  ``up movement" at the time $t=1$ $Y^0_{1,1}$ changes sign: $Y^0_{1,1} = -1$.

After the second ``down movement", $X^1$ becomes minus $Y^1$, $X^1_{1,1} = 2$, and
after the second ``up movement" at the time $t=2$  $Y^0_{1,1}$ changes sign again: $Y^0_{1,1} = 1$, etc.

Here we have obtained a simple oscillation of a network weight, $Y^0_{1,1}$
(the network matrix is $Y^0$ at any given moment of time).

\subsubsection{The {\tt aug\_27\_16\_experiment} directory} 

Here instead of $Y^1$ we
take a collection of constant update matrices, $Y^{j_1}, \dots Y^{j_n}$. Just like in
the previous example, make sure that the first rows (indexed by 0) of those matrices are 0.
For the second rows (indexed by 1), take $Y^{j_1}_{1,j_1} = -1, Y^{j_1}_{1,j_2} = 1,\newline
Y^{j_2}_{1,j_2} = -1, Y^{j_2}_{1,j_3} = 1, \ldots,
Y^{j_{n-1}}_{1,j_{n-1}} = -1, Y^{j_{n-1}}_{1,j_n} = 1,\newline
Y^{j_n}_{1,j_n} = -1, Y^{j_n}_{1,j_1} = 1$, and the rest of the elements of
the second rows of these matrices are 0.

Start at $t=0$ with $Y^0$ matrix having the first row as before, and the second row containing
the only non-zero element $Y^0_{1,j_1} = 1$. Then one can easily see (or verify by
downloading and running under Processing 2.2.1 the open source software in the {\tt Lightweight\_Pure\_DMMs/aug\_27\_16\_experiment}
directory of the Project Fluid~\cite{Fluid}) that at the moment $t=1$ the only non-zero
element in the second row of $Y^0$ is $Y^0_{1,j_2} = 1$, at the moment
$t=2$ the only non-zero
element in the second row of $Y^0$ is $Y^0_{1,j_3} = 1$, and so on until at the moment
$t=n$ this wave of network connectivity pattern loops back to $Y^0_{1,j_1} = 1$, and then continues
looping indefinitely through these $n$ states\footnote{{\bf July 2018 note:} see 
\url{https://arxiv.org/abs/1706.00648} (Appendix B) for a more polished implementation and presentation}.

\subsubsection{Final remarks} 

\paragraph{A.} The actual implementation of {\tt Self} in the prototype enforces the constraint that  $Y^0_{0,0} = 1$, $Y^0_{0,j} = 0$ for all $j \neq 0$.

\paragraph{B.} By making the update matrices dynamically dependent upon e.g. input symbols, one could embed an arbitrary
deterministic finite automaton into this control mechanism in this fashion.

\section{Computation with Involutions}\label{sec:involutions}

Recall from Appendix~\ref{sec:rectifiers_and_quasi_metrics} that {\em rectified linear unit} (ReLU) is a neuron with the activation function $f(x) = \mbox{max}(0,x)$.

Putting together Appendices~\ref{sec:interval_monotonic} and~\ref{sec:rectifiers_and_quasi_metrics} we obtain
$$x \sqsubseteq  (x+[f(a_1), -f(-b_1)]) \sqsubseteq$$ $$ (x+[f(a_1), -f(-b_1)]+[f(a_2), -f(-b_2)]) \sqsubseteq \dots$$

Here negations before and after application of ReLU in the terms $-f(-b_1), -f(-b_2), \dots$ are anti-monotonic involutions.
Since we apply them twice, the overall inference remains monotonic.
(Cf. the use of anti-monotonic involutions
to perform anti-monotonic inference in Section 4.14 ``Computational Models with Involutions" of~\cite{MBukatinMatthewsLinear}.)\footnote{The material 
in the Appendix~\ref{sec:involutions} was added in July 2018. For the first appearance of this material see slide 18 of 
{\scriptsize \url{http://www.cs.brandeis.edu/~bukatin/DMMsPrefixTreesMar2017.pdf}}.
 Another pattern similar in spirit (slide 19 of the same slide deck) is {\bf Concatenated ReLU},
 $x \mapsto (f(x), f(-x))$, introduced in [Wenling Shang, Kihyuk Sohn, Diogo Almeida, and Honglak Leel, 
{\em Understanding and Improving Convolutional Neural Networks via Concatenated Rectified Linear Units}, 
2016, \url{https://arxiv.org/abs/1603.05201}] to address the ``problem of dying ReLUs".
We can think about this as incorporating both (dual to each other) quasi-metrics on the reals mentioned in Appendix~\ref{sec:rectifiers_and_quasi_metrics}.
In terms of scalar neurons, this is a neuron with two outputs.}

\section{Dataflow Matrix Machines Based on Streams of V-values and Variadic Neurons versus Lightweight Pure Dataflow Matrix Machines}\label{sec:variadic_vs_pure}

V-values are vector-like elements based on nested maps. They subsume pure dataflow matrix machines in terms of their
ability to have sufficiently expressive dataflow matrix machines based on a single kind of linear streams. In addition,
they support variadic neurons, so there is no need to explicitly keep track of input and output arities of neurons.
V-values conveniently represent a variety of conventional data structures, so the intricate machinery of Section~\ref{as_matrices}
of the present paper is unnecessary for DMMs based on streams of V-values.

However, there is one aspect where, in particular, {\em lightweight} pure dataflow machines still
have considerable advantage at the moment. Lightweight Pure DMMs have highly regular structure
which is friendly for batching and for GPUs. In contrast, DMMs based on V-values and variadic neurons
often have highly irregular structure, which is also allowed to vary with time. 

Making DMMs based on V-values and variadic neurons friendly for batching and for GPUs is, at
present, an open problem (although such advances as {\em dynamic batching} underlying 
TensorFlow Fold library is a good indication that this problem will eventually be solved for DMMs based on 
V-values and variadic neurons as well).\footnote{The material 
in the Appendix~\ref{sec:variadic_vs_pure} was added in July 2018. DMMs based on streams of V-values
and variadic neurons are described in [Michael Bukatin and Jon Anthony,
{\em Dataflow Matrix Machines and V-values: a Bridge between Programs and Neural Nets}, 2017,
\url{https://arxiv.org/abs/1712.07447}]. The dynamic batching used in TensorFlow Fold
library is introduced in [Moshe Looks, Marcello Herreshoff, DeLesley Hutchins, and Peter Norvig, 
{\em Deep Learning with Dynamic Computation Graphs}, 2017, \url{https://arxiv.org/abs/1702.02181}].}

Further experiments with self-referential DMMs were performed by participants of DMM and Fluid projects in January-October 2018.  In particular, 
self-referential facilities of DMMs based on V-values and variadic neurons were used to interactively edit a running network
on the fly ({\bf ``livecoding"}) and a series of experiments with randomly initialized Lightweight Pure DMMs produced emerging bistable behavior in a variety of different settings.\footnote{See
Section 1 of [DMM technical report 11-2018, {\em Dataflow matrix machines: recent experiments and notes for next steps}, Preprint, Nov. 2018,
{\tiny \url{https://github.com/jsa-aerial/DMM/blob/master/technical-report-2018/dmm-notes-2018.pdf}}].}

\end{document}